\documentclass[prl,aps,showpacs,12pt,reprint,superscriptaddress,longbibliography]{revtex4-1}
\usepackage[utf8]{inputenc}

\usepackage{graphicx}
\usepackage{amsmath}
\usepackage{amsfonts}
\usepackage{amssymb,mathbbol,yfonts}
\usepackage{color}					
\usepackage{bm}      				
\usepackage{booktabs}				
\usepackage{array}					
\usepackage{slashed}  				
\usepackage[hidelinks]{hyperref}	
\hypersetup{colorlinks=true,citecolor=blue,urlcolor=blue} 
\usepackage{tikz}					
\usepackage{verbatim}

\newcommand{\MF}{\textsc{mf}}

\newcommand{\bg}{\bm{\gamma}}

\renewcommand{\l}{\mathrm{l}}
\newcommand{\bea}{\begin{eqnarray}}
\newcommand{\eea}{\end{eqnarray}}
\newcommand{\beq}{\begin{equation}}
\newcommand{\eeq}{\end{equation}}
\newcommand{\bit}{\begin{itemize}}
\newcommand{\eit}{\end{itemize}}

\begin{document}

\title{Efficiency Fluctuations of Stochastic Machines Undergoing a Phase Transition}

\author{Hadrien Vroylandt}
\affiliation{Universit\'e Paris-Saclay, CNRS/IN2P3, IJCLab, 91405 Orsay, France}
\author{Massimiliano Esposito}
\affiliation{Complex Systems and Statistical Mechanics, Department of Physics and Material Science, University of Luxembourg, L-1511 Luxembourg, G.D. Luxembourg}
\author{Gatien Verley}
\affiliation{Universit\'e Paris-Saclay, CNRS/IN2P3, IJCLab, 91405 Orsay, France}
\date{\today}

\begin{abstract}
We study the efficiency fluctuations of a stochastic heat engine made of $N$ interacting unicyclic machines and undergoing a phase transition in the macroscopic limit. Depending on $N$ and on the observation time, the machine can explore its whole phase space or not. This affects the engine efficiency that either strongly fluctuates on a large interval of equiprobable efficiencies (ergodic case) or fluctuates close to several most likely values (nonergodic case). We also provide a proof that despite the phase transition, the decay rate of the efficiency distribution at the reversible efficiency remains largest one although other efficiencies can now decay equally fast. 
\end{abstract}

\maketitle

\textit{Introduction.---} Small machines behave on average like macroscopic ones: a mean input flux is converted into a mean output flux with an efficiency bounded by the reversible efficiency due to the second law of thermodynamics \cite{Callen1985_vol}. However, their input and output fluxes fluctuate with root mean squares which can be larger than their averages. These fluctuations are constrained by the universal fluctuation relations that lead to the second law at the ensemble averaged level~\cite{Sinitsyn2011_vol44, Campisi2014_vol47, Rao2018_vol20}. 
This implies that the efficiency $\eta$ of the machine along a single realization of duration $t$ is also a stochastic quantity characterized by a probability distribution $P(\eta)$. 
As recently discovered, its fluctuations also display universal statistical features in both classical~\cite{Verley2014_vol90, Gingrich2014_vol16, Vroylandt2016_vol93, Verley2014_vol5, Polettini2015_vol114, Proesmans2015_vol17, Proesmans2015_vol109, Proesmans2015_vol92} and quantum systems \cite{Esposito2015_vol91, Jiang2015_vol115, Agarwalla2015_vol92, Denzler2019vol}. 
More specifically, for long trajectories of autonomous machines, the distribution $P(\eta)$ concentrates at the macroscopic efficiency $\bar \eta$ while the reversible efficiency $\eta_\mathrm{rev}$ becomes asymptotically the less likely. Also, the efficiency large deviation function (LDF), defined as the long time limit of $t^{-1} \ln P(\eta)$, has a characteristic smooth form with two extrema only and a well-defined limit for large efficiency fluctuations. 
These predictions were experimentally verified in Refs.\cite{Martinez2015_vol, Proesmans2016_vol6}.
However, these results focus on the efficiency statistics at long times and rely on the assumptions that the machine has a finite state space and thus cannot undergo a phase transition. 

The performance of machines undergoing a nonequilibrium phase transition has attracted increasing attention \cite{Imparato2015_vol17,Golubeva2012_vol109,Golubeva2014_vol89,Campisi2016_vol7,Pyoung2016_PRE,Herpich2018PRX,Herpich2019PRE}.
In this Letter, we consider a model of $N$ interacting machines first proposed in Ref.~\cite{Cleuren2001_vol54}.
At the mean-field (MF) level, i.e. when $N \to \infty$, they may undergo a nonequilibrium phase transition caused by an asymmetric pitchfork bifurcation. 
Past the bifurcation point, ergodicity is broken and these machines exhibit multiple macroscopic efficiencies~\cite{Vroylandt2017_vol120}. 
In practice this means that their initial condition will determine which stable steady state is eventually reached and its corresponding macroscopic efficiency. 
As a result fluctuations in performance only come from uncertainties in the initial state. 
Our main goal here is to characterize how efficiency fluctuations scale in size $N$ and in time $t$ in such critical machines using LDFs. 
We do so by developing a path integral method (in the spirit of \cite{Book_Ross2008, Ritort2004_vol2004, Tailleur2008_vol41, Grafke2017_vol2017, Suarez1995_vol102, Weber2017_vol80, Lazarescu2019vol151}). 
Crucially two regimes must be distinguished depending on the order in which these scalings are taken, each yielding to a different LDF.
The first, $J(\eta)$, characterizes the nonergodic regime and corresponds to taking first $N \to \infty$ and then $t \to \infty$ on $(Nt)^{-1} \ln P(\eta)$. 
The second, $J^{**}(\eta)$, characterizes the ergodic regime and corresponds to the opposite order of limits.
While this latter remains smooth, its two extrema become degenerate, giving rise to strong efficiency fluctuations spanning over different operating modes.
The former instead is not continuously differentiable anymore and displays steep minima located around the mean field efficiencies and multiple plateaux. 
Remarkably, despite significant qualitative changes in both types of LDF, the reversible efficiency, while not uniquely anymore, has the fastest decaying efficiency probability. 
While our method is presented for a specific model, it seems particularly well suited to study collections of interacting machines and characterizes critical nonequilibrium fluctuations.   

\begin{figure}[h]
\centering
\includegraphics[width=\columnwidth]{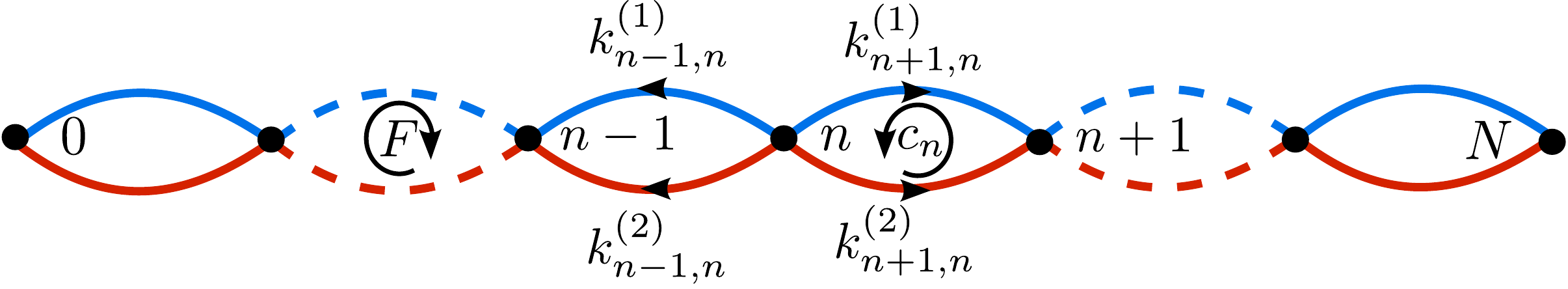}
\caption{\label{graph} 
Graph of the discrete state space of the collective machine. Blue edges are for channel $1$ and red edges for channel $2$.
}
\end{figure}

\textit{Model ---}
We consider a machine made of a collection of $N$ interacting unicyclic machines. Each of these is autonomous and converts heat into mechanical work by hopping between two discrete states of energy $0$ or $E\geq 0$ via two different transition channels labeled by $\nu$, where $\nu = 1$ is caused by a cold reservoir at temperature $T^{(1)} = 1/\beta^{(1)}$ and $\nu = 2$ by a hot one at $T^{(2)}=1/\beta^{(2)}$ (we set $k_{B}=1$). A nonconservative force promotes (resp. represses) the transition from the lower to the higher energy state via channel $\nu = 1$ (resp. $\nu =2$), while the opposite is true for the transition from the higher to the lower state. These unicyclic machines interacts via a pair interaction energy $V/N$ only when they are not in the same states. The energy of the collective machine thus reads $U_{n} = nE+ n (N-n)V/N$, where $n$ is the number of machines in the high energy state. The probability to find the collective machine in state $n$ at time $t$ follows a Markov master equation $\dot p_n = \sum_{\epsilon = \pm 1, 0} \sum_{\nu} k^{(\nu)}_{n,n+\epsilon} p_{n+\epsilon}$, where $k^{(\nu)}_{n+\epsilon,n}$ is the Poisson rate with which a unicyclic machine hops to a high (resp. low) energy state for $\epsilon = 1$ (resp. $\epsilon = -1$) via channel~$\nu$ and $k^{(\nu)}_{n,n} = - k^{(\nu)}_{n+1,n}-k^{(\nu)}_{n-1,n}$, see Fig.~\ref{graph}. 
To specify further the dynamics, we choose (for $\epsilon = \pm 1$)
\begin{equation}
k^{(\nu)}_{n+\epsilon,n} = N \left( \frac{1+\epsilon}{2}-\epsilon \frac{n}{N} \right)e^{ -\frac{\beta^{(\nu)}}{2} \left(E_{a} + U_{n+\epsilon}- U_{n} - W^{(\nu)}_{n+\epsilon,n} \right)  } \label{eq:LDB},
\end{equation}
where $E_{a}$ is an activation energy and $W^{(\nu)}_{n+\epsilon,n} \equiv -\epsilon(-1)^\nu F$ is the work done by the nonconservative force and received by the machine during the transition $n \rightarrow n+\epsilon$ via $\nu$.
Defining intensive quantities as being per unicyclic machine and per unit time, the intensive stochastic heat from the hot reservoir and the intensive work from the nonconservative force are, respectively 
\begin{equation} 
q = \sum_{n=0}^{N-1} \phi_{\mathrm{q},n} j^{(2)}_{n} \quad \text{and} \quad w=\sum_{n=0}^{N-1}\phi_{\mathrm{w},n} j^{(2)}_{n},
\end{equation}
where $j^{(2)}_{n}$ counts the intensive net number of jumps from $n$ to $n+1$ via channel $2$ in a stochastic trajectory. 
Indeed, when $X=\mathrm{q}$ (resp. $X=\mathrm{w}$), $\phi_{X,n} $ gives the amount of energy received from the hot reservoir (resp. from the nonconservative force) when the system undergoes a cycle $c_{n} \equiv \left(n \underset{\nu = 2}{\longrightarrow} n+1 \underset{\nu = 1}{\longrightarrow } n \right) $:
\begin{eqnarray}
	\phi_{\mathrm{q},n} & \equiv & U_{n+1}-U_{n}-W^{(2)}_{n+1,n} \simeq V(1-2n/N)+F, \\
	\phi_{\mathrm{w},n} & \equiv & W^{(1)}_{n,n+1}+W^{(2)}_{n+1,n} = -2F.
\end{eqnarray}
The intensive stochastic entropy production $\sigma \equiv \sigma^\mathrm{w} + \sigma^\mathrm{q}$ is the sum of the two partial entropy production $\sigma^\mathrm{q} \equiv \left[\beta^{(1)}-\beta^{(2)} \right] q$ and $\sigma^\mathrm{w} \equiv \beta^{(1)}w$. The stochastic efficiency is thus defined as $\eta \equiv-\sigma^\mathrm{w}/\sigma^\mathrm{q}$. 
Their local (i.e. along each cycle $c_{n}$) analogs read $\sigma^\mathrm{q}_{n} \equiv \left[\beta^{(1)}-\beta^{(2)} \right]\phi_{\mathrm{q},n}$, $\sigma^\mathrm{w}_{n} \equiv \beta^{(1)} \phi_{\mathrm{w},n} $,  
\begin{equation}
\label{eq:localeff}
	\eta_{n}^\l \equiv -\frac{\beta^{(1)} \phi_{\mathrm{w},n}}{\left( \beta^{(1)}-\beta^{(2)}\right) \phi_{\mathrm{q},n}} = - \frac{\sigma^\mathrm{w}_{n}}{\sigma^\mathrm{q}_{n}} . 
\end{equation} 
In the macroscopic limit where $N$ is very large and the density of units in the high energy state $x=n/N$ can be treated as a continuous variable, we denote them, respectively, by $\sigma^\mathrm{q}_{x}$, $\sigma^\mathrm{w}_{x}$ and $\eta^{l}_{x}$.

\begin{figure}[t]
\centering
\includegraphics[width=\columnwidth]{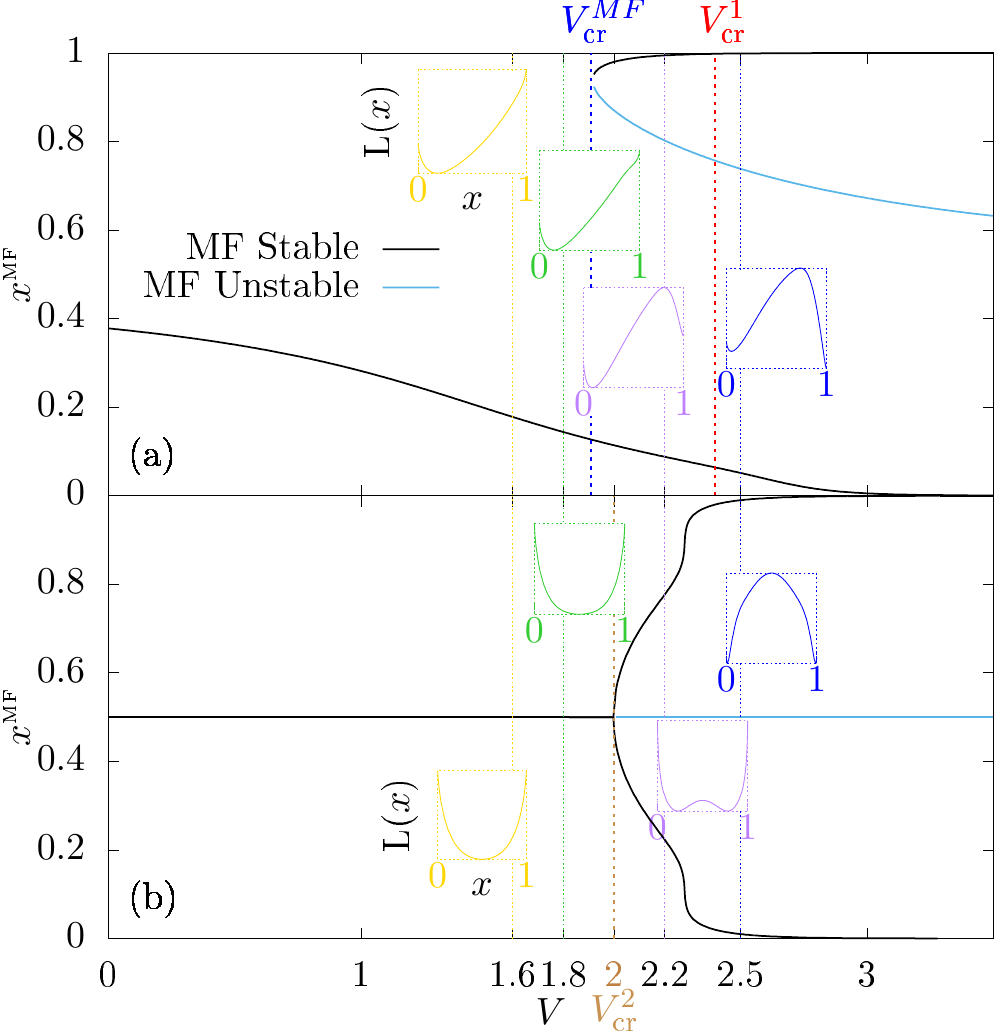}
\caption{\label{FigMF} 
Stable (black) and unstable (light blue) mean field steady state densities $x^\MF$ versus interaction energy $V$. 
Insets: density LDFs versus $x$ for four values of $V$ indicated by vertical dashed lines. 
The parameters are $E_a=2$, $E=0$, $\beta^{(1)}=10$, $F=0.5$ for panel (a) 
and $F=0$ for panel (b).
In all the Letter we take $\beta^{(2)}=1$ to set the energy scales, while the timescale is set by Eq.~(\ref{eq:LDB}).
}
\end{figure}

\textit{Mean field dynamics ---} When $N \to \infty$ but $t$ remains finite, the master equation becomes a nonlinear MF master equation for $x$ \cite{Cleuren2001_vol54}. Ergodicity breaking is evidenced by the fact that its stationary solutions may take one, three (or even five) values $x^\MF$ depending on $V$ and $F$, as shown on the branching diagrams of Fig.~\ref{FigMF}. Each of these solutions will give rise to a corresponding MF efficiency trough Eq.~\eqref{eq:localeff}. The MF master equation is exact for this model, i.e., the extrema of the density LDF $L(x)$ \cite{Book_Hill1989,Vroylandt2017_vol120} (shown in the insets) coincide with the MF densities. In panel (a) for $F=0.5$, the abrupt change in the position of the minimum of the density LDFs around $V_\mathrm{cr}^{1}$ reveals a first order phase transition while in panel (b) for $F=0$ the smooth appearance of two minima at $V_\mathrm{cr}^{2}$ reveals a second order phase transition. 

\textit{Currents and efficiency fluctuations ---}
The quantity of interest is the cumulant generating function (CGF) for $\sigma^\mathrm{q}$ and $\sigma^\mathrm{w}$ expressed in terms of their conjugated Laplace parameter $\bg = (\gamma^\mathrm{q},\gamma^\mathrm{w})$ which reads
\begin{equation}
\Phi(\bm{\gamma}) \equiv \lim_{Nt \to \infty}\frac{1}{Nt} \ln \left \langle  e^{N t (\gamma^\mathrm{q} \sigma^\mathrm{q} +\gamma^\mathrm{w} \sigma^\mathrm{w} )} \right \rangle_{p_{0}}  \label{MomentGeneratingFunction},
\end{equation}
where $\left\langle \dots \right\rangle_{p_{0}}$ is the mean on paths with initial condition drawn from probability density $p_{0}$.
Using path integral technique \cite{Vroylandt2018_vol2018,Lazarescu2019vol151,Phdthesis_Vroylandt2018}, this CGF can be written as the maximum value taken by an action over trajectories $[x]_{0}^{\infty}$ of infinite duration
\begin{equation}
\label{eq:CGFfromAction}
	\Phi(\bm{\gamma}) =  \max_{[x]_{0}^{\infty}} S([x]_{0}^{\infty},\bm{\gamma}).
\end{equation}
The action $S([x]_{0}^{t},\bm{\gamma}) = (1/t)\int_{0}^{t} \mathrm{d}\tau  \mathcal{L}(x(\tau),\dot x(\tau),\bm{\gamma})$  is associated to the Lagrangian given by
\begin{align}
  \label{eq:Lagrangien_Donkey} \nonumber
  \!\!\!\mathcal{L}(x,\dot x,\bm{\gamma}) \equiv 
    \sqrt{\dot{x}^2+\varphi(x,\bm{\gamma})}-\sum_{\epsilon=\pm1,\nu=1,2} J_{\epsilon;\nu}(x)
 \\ + \dot{x}\ln\left( \frac{- \dot x + \sqrt{\dot{x}^2+\varphi(x,\bm{\gamma})}}{2\sum_{\nu=1,2} J_{-1;\nu}(x) e^{- (\gamma^\mathrm{q}\sigma^\mathrm{q}_{x}+\gamma^\mathrm{w}\sigma^\mathrm{w}_{x}) \delta_{\nu,2} }} \right),
\end{align}
where we introduced the transition rates in the continuous limit $J_{\epsilon;\nu} \equiv \lim_{N\rightarrow \infty} k_{xN+\epsilon,xN}^{(\nu)}/N$ and the function
\begin{equation}
\varphi(x,\bm{\gamma}) \equiv 4\prod\limits_{\epsilon=\pm1}\sum_{\nu=1,2} J_{\epsilon;\nu}(x) \exp{ \left[ \epsilon (\gamma^\mathrm{q}\sigma^\mathrm{q}_{x}+\gamma^\mathrm{w}\sigma^\mathrm{w}_{x}) \delta_{\nu,2} \right]}.
\end{equation}
From extremum action principle, $\Phi(\bm{\gamma})$ is the action evaluated for the optimal trajectories satisfying the Euler-Lagrange equation based on Lagrangian \eqref{eq:Lagrangien_Donkey} for given initial conditions $x(0)$ and $\dot x(0)$. The remaining optimization on initial conditions amounts to select stationary trajectories only since the CGF is bounded by 
\begin{equation}
	\max_{stat. [x]} S[x] \leq \Phi(\bm{\gamma}) \leq \max_{x,\dot x} \mathcal{L}(x,\dot x, \bm{\gamma}).
\end{equation}
The lower bound arises from restricting the maximization to the subset of stationary trajectories (i.e. trajectories with constant density), while the upper bound follows from exchanging the maximization and the time integration in the action. 
For Lagrangian~\eqref{eq:Lagrangien_Donkey}, the maxima in the upper bound can be shown to coincide with the stationary solutions $x^{*}$ of Euler-Lagrange equation. 
Hence, the upper and lower bounds match yielding the CGF 
\begin{equation} 
\Phi(\bm{\gamma}) = \max_{x^{*}} \mathcal{L} (x^*,0,\bm{\gamma})  \label{eq:propaGF_Donkey}.
\end{equation} 
The LDF for stochastic efficiency can be computed from the CGF of the partial entropy productions directly \cite{Verley2014_vol90}.
When $x^{*}$ is not unique, the order of the limits $t\rightarrow \infty$ and $N\rightarrow \infty$ in (\ref{MomentGeneratingFunction}) is of importance \cite{Vroylandt2019_vol174}. 
In the \emph{ergodic case}, the initial probability density $p_{0}$ plays no role and the $x^{*}$ maximizing the value of the Lagrangian is chosen in Eq.~\ref{eq:propaGF_Donkey}. The efficiency LDF then reads
\begin{eqnarray}
J^{**}(\eta) & \equiv & - \min_{\gamma^\mathrm{w}} \max_{x^{*}} \mathcal{L} (x^{*},0,\gamma^\mathrm{w}\eta,\gamma^\mathrm{w}) \label{eq:Jergo} \\ &=& - \min_{\gamma^\mathrm{w}}\Phi(\gamma^\mathrm{w}\eta,\gamma^\mathrm{w}) \geq 0,  \label{standardCGFtoJ}
\end{eqnarray}
where we used $\Phi(0,0)=0$.
In the \emph{nonergodic case}, the system can be separated into ergodic regions and the number of regions accessible with the chosen initial condition $p_{0}$ will matter \cite{Vroylandt2019_vol174}. The $x^{*}$ which belongs to those accessible regions and which maximizes the value of the Lagrangian must be picked. The efficiency LDF reads
\begin{equation}
J(\eta) \equiv - \max_{x^{*}}\min_{\gamma^\mathrm{w}} \mathcal{L} (x^{*},0,\gamma^\mathrm{w}\eta,\gamma^\mathrm{w})  \label{eq:Jnonergo},
\end{equation} 
where the maximum holds on all $x^{*}$ when choosing a uniform initial condition that makes all ergodic regions accessible.

\textit{Results.---} 
The signature of a phase transition and/or ergodicity breaking is when $x^*$ stops being unique. While the CGF is always continuous and convex, its derivatives may become singular \cite{[{}][{, Section 3.5.2. }]Touchette2009_vol478}. A kink in the CGF signals a nonconvexity or a linear part in the currents LDF.
We now proceed to prove that the reversible efficiency still corresponds to the faster decay rate of the efficiency probability without using the convexity of the LDF. 
The fluctuation relation $\Phi(\bm{\gamma}) = \Phi(-\bm{\gamma} -\bm{1})$ imposes that $\Phi$ is symmetric with respect to the point $\bm{\gamma} = (-1/2,-1/2)$ which we denote by C. Then, since $\Phi$ is convex, it has a minimum at C and the minima of $\mathcal{L}$ in Eqs.~(\ref{eq:Jergo}) or (\ref{eq:Jnonergo}) are reached at this point when the efficiency is the reversible one ($\eta = 1$) leading to $J(\eta) \leq J(1)$. However, since $\Phi$ is not necessarily strictly convex, the minima may be degenerate and other efficiencies can give rise to equally large LDF. 

\begin{figure*}
	\begin{center}
		\includegraphics[width=\linewidth]{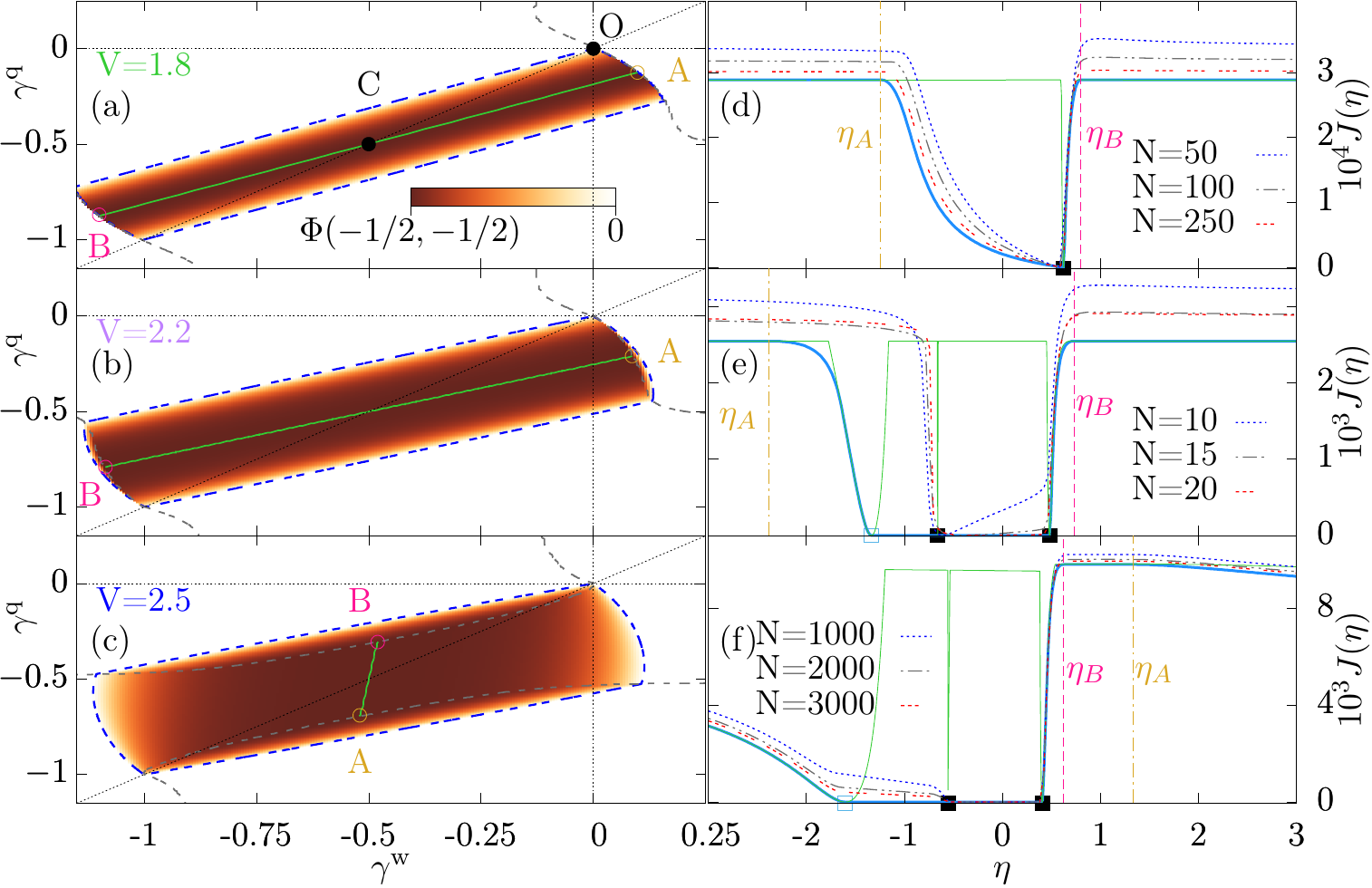}
\caption{
Left column, CGFs $\Phi$ defined in \eqref{eq:propaGF_Donkey} as a function of $\gamma^\mathrm{q}$ and $\gamma^\mathrm{w}$ for three different $V$ (a--c). Right column, the corresponding efficiency LDF (d--f). 
On the left, the diagonal black dotted line of slope one is there to guide the eye and the blue dashed line is the contour line $\Phi = 0$ enclosing the $\Phi < 0$ region that is relevant to calculate \eqref{standardCGFtoJ}. The green solid line between A and B defines the degenerate minimum of the CGF. Its boundaries belong to the dashed gray critical lines separating regions with different dominant stationary solutions $x^{*}$. The slopes $\eta_A$ and $\eta_B$, of the lines (OA) and (OB) respectively, give the efficiencies delimiting the higher plateaux of the efficiency LDFs on the right.
When $V>V_\mathrm{cr}^\MF=1.92$, one critical line touches the origin indicating bistability in the MF dynamics.
On the right, $J^{**}(\eta)$, $J(\eta)$ and the finite $N$ efficiency LDFs are given, respectively, by the light blue thick solid line, the thin green solid line and the different dashed lines. The solid black (resp. empty blue) squares show the location of the stable (resp. unstable) MF efficiencies. The parameters are those of Fig.~\ref{FigMF}(a).  \label{fig:JmicroEtCano} 
}
	\end{center}
\end{figure*}

We now turn to our numerical results. In Figs.~\ref{fig:JmicroEtCano}d--f, we show the efficiency LDFs obtained from Eqs.~(\ref{eq:Jergo}--\ref{eq:Jnonergo}) (for $N \to \infty$) or from numerical evaluation of the CGF for $\sigma^\mathrm{q}$ and $\sigma^\mathrm{w}$ (for finite $N$) using standard spectral techniques \cite{Chetrite2015_vol16,Esposito2007_vol76}. 
We clearly see that both $J(\eta)$ and $J^{**}(\eta)$ are substantially different than the efficiency LDF of finite machines discussed in Ref.~\cite{Verley2014_vol90}.
In both cases their maximum is degenerate and comprises the reversible efficiency as we will explain below. We remark that $J(\eta)>J^{**}(\eta)$ for all $\eta$, as expected since $J^{**}(\eta)$ can be derived from the convex hull of the nonconvex LDF for partial entropy productions from which $J(\eta)$ is derived \cite{Vroylandt2019_vol174}.   
The minimum of both LDF that correspond to the MF efficiency is unique for $V<V_\mathrm{cr}^\MF$, while for higher $V$, a plateau connects the different MF efficiencies $\eta^\mathrm{l}_{x^{\MF}}$ in the ergodic case or several minima appear in $J(\eta)$ in the nonergodic case. The plateaux signify that ergodicity enables large fluctuations between MF efficiencies while nonergodicity prevents them. Interestingly, our numerical computations for increasing $N$ show a faster convergence of $J^{**}(\eta)$ toward the plateau lying between two \emph{stable} MF efficiencies. 
Efficiency LDFs with multiple minima (or even a plateau) had not been reported before. Finding these plateaux and relating them to the existence of a phase transition in the machine constitutes a key finding of this Letter.  

We now discuss the physical origin of the degenerate maximum of efficiency LDF. In tightly coupled finite machines \cite{Esposito2009_vol102,Cleuren2015_vol224}, the input and output fluxes are proportional at the stochastic trajectory level ($-\sigma^\mathrm{q} \alpha = \sigma^\mathrm{w}$) and $\bar \eta=\alpha$. As a result, the CGF $\Phi(\bm{\gamma})$ displays a translation invariance: it is zero on the line $\gamma^\mathrm{q}-\alpha \gamma^\mathrm{w} = 0$ and constant on any other parallel line. Using (\ref{standardCGFtoJ}), the efficiency LDF has a singular minimum zero at efficiency $\alpha$ and a degenerate maximum everywhere else \cite{Polettini2015_vol114}.  
This results from the fact that the stochastic efficiency is either a constant number $\alpha$ or undefined when both $\sigma^\mathrm{q}$ and $\sigma^\mathrm{w}$ are zero (or more precisely subextensive in $Nt$). The degenerate LDF value thus corresponds to the LDF of the probability of having no extensive hot heat input and work output.
However, when such machines have infinite state spaces, the notion of tight coupling softens as extensive entropy fluctuations can arise and compromises the translation invariance of the CGF (in fact it remains valid in a bounded region and the CGF diverges elsewhere). As a result the efficiency LDF still displays a degenerate maximum but that does not cover anymore all the efficiencies since the minimum is not singular anymore and is reached continuously \cite{Park2016_vol94, Manikandan2019vol122}. 
In our model, similar plateaux are observed in Figs.~\ref{fig:JmicroEtCano}d--f. However the mechanism responsible for softening the tight coupling is different and is the phase transition. 
The CGF has no global translation invariance anymore, but the Lagrangian keeps some invariance upon change of $\bm{\gamma}$ as one can check directly 
\begin{equation}
    \mathcal{L}\left(x,0,-\frac{1}{2}+\left[\gamma^\mathrm{w}+\frac{1}{2} \right] \eta^{l}_{x},\gamma^\mathrm{w}\right) = \mathcal{L}\left(x,0,-\frac{1}{2},-\frac{1}{2}\right). \label{symmetry}
\end{equation}
For each density $x^*$ over which the maximization is taken in \eqref{eq:Jergo} and \eqref{eq:Jnonergo} and for given $\eta \neq \eta^\mathrm{l}_{x^*}$, the Lagrangian minimizer $\bm{\gamma}^\mathrm{w} = (\eta^\mathrm{l}_{x^*}-1)/(2\eta-2\eta^\mathrm{l}_{x^*})$ is yielding the same minimum $\mathcal{L}(x^*,0,-1/2,-1/2)$ as long as the phase transition induces no change of maximizer $x^*$ (this happens at efficiency $\eta_A$ and $\eta_B$). This degeneracy is illustrated for the absolute minimum $\Phi(-\frac{1}{2},-\frac{1}{2})$ on Figs.~\ref{fig:JmicroEtCano}a--c. In the end, several $\eta$'s share the same Lagrangian's minimum associated to the same maximum $J(1)$ of the efficiency LDF in both the ergodic and nonergodic cases. As in tightly coupled finite machines, these degenerate LDF maxima correspond to the LDF of the probability for no extensive work and hot heat to arise.

In summary, using a simple model, we found that efficiency fluctuations are strongly affected by the existence of a phase transition and depend on the order in which the long time and large size limit are taken. Nonetheless, the efficiency probability still decay the faster at the reversible efficiency, but maybe decay equally fast at other efficiencies. Our large deviation theory techniques are general and opens the way to a more systematic study of efficiency fluctuations in energy converters undergoing a phase transition.

\begin{acknowledgments}
	We dedicate this work to Christian Van den Broeck who initiated this research project. We thank Alexandre Lazarescu for his comments on path action extremization. M. E. is funded by the European Research Council project NanoThermo (ERC-2015-CoG Agreement No. 681456).
\end{acknowledgments}

\bibliography{Ma_base_de_papier}

\end{document}